\documentclass{article}

\usepackage{graphicx} 
\usepackage{geometry}
\newgeometry{vmargin={20mm}, hmargin={30mm,30mm}} 
\usepackage[most]{tcolorbox}
\usepackage[HTML]{xcolor}
\usepackage{minted}
\usepackage{hyperref}
\usepackage{titling} 
\usepackage{booktabs}
\usepackage{enumitem}
\usepackage[normalem]{ulem}  
\usepackage{subcaption} 
\usepackage{array}
\usepackage{longtable}
\usepackage{algorithm}
\usepackage{algpseudocode}
\usepackage{float}

\usepackage{newtxtext,newtxmath} 

\usepackage{setspace}
\newcounter{roundedboxctr}

\newtcolorbox{roundedbox}[2][]{%
  width=0.95\textwidth,
  colback=white,
  colframe=gray!89,
  fonttitle=\bfseries,
  arc=1.4mm,
  boxrule=0.2pt,
  enhanced,
  boxed title style={colback=gray!10, colframe=gray!10},
  title={%
    Box~\theroundedboxctr: #2%
  },
  #1
}

\pretitle{%
  \vspace*{-1em}
\centering
  \rule{\textwidth}{1.5pt}\par\vspace{0.7em}%
  \LARGE\bfseries
}

\posttitle{%
  \par
  \rule{\textwidth}{0.8pt} 
}

\preauthor{\par\vspace{0em}\centering}
\postauthor{\par\vspace{1.5em}}
\predate{\par\centering\large}
\postdate{\par}

\newcommand{\AuthorBlockNew}[2]{%
  \begin{tabular}[t]{@{}c@{}}
    \textbf{#1}\\
    #2
  \end{tabular}%
}

\title{Stacked Regression using Off-the-shelf, Stimulus-tuned and Fine-tuned Neural Networks for Predicting fMRI Brain Responses to Movies (Algonauts 2025 Report)}

\author{%
\centering
\AuthorBlockNew{Robert Scholz\textsuperscript{1,2*}}{Université Paris Cité} \quad \AuthorBlockNew{Kunal Bagga\textsuperscript{*}}{Indep. Researcher} \quad \AuthorBlockNew{Christine Ahrends\textsuperscript{}}{University of Oxford} \quad \AuthorBlockNew{Carlo Alberto Barbano\textsuperscript{}}{University of Turin} \\[1em] {\centering
\textsuperscript{1} Universität Leipzig, \textsuperscript{2} Max Planck School of Cognition \\
\textsuperscript{*}These authors contributed equally and can be reached via \\\ robert.scholz [at] maxplanckschools.de, kunal [at] kunalb.com
}
}

\date{}

\begin{document}

\maketitle

\begin{abstract}

We present our submission to the Algonauts 2025 Challenge, where the goal is to predict fMRI brain responses to movie stimuli. Our approach integrates multimodal representations from large language models, video encoders, audio models, and vision–language models, combining both off-the-shelf and fine-tuned variants. To improve performance, we enhanced textual inputs with detailed transcripts and summaries, and we explored stimulus-tuning and fine-tuning strategies for language and vision models. Predictions from individual models were combined using stacked regression, yielding solid results. Our submission, under the team name Seinfeld, ranked 10th. We make all code and resources publicly available, contributing to ongoing efforts in developing multimodal encoding models for brain activity.

\end{abstract}

\section{Introduction}

Encoding models predict brain responses to a set of given stimuli. Recently, deep neural networks have been used as encoding models to predict brain activity as recorded by functional MRI (fMRI) \cite{kell2018task, jain2018incorporating, LeBel2021, caucheteux2022deep, doerig2025high, oota2023deep}. These studies investigate whether representations in deep neural networks correspond to those in the human brain. This relationship is often assessed using linear models, with successful prediction taken as evidence of shared representational structure. Studies have investigated representations from both unimodal and multimodal deep neural networks, including large language models (LLMs) \cite{jain2018incorporating, caucheteux2022deep, antonello2023scaling, tuckute2024driving}, vision models \cite{eickenberg2017seeing, Guclu_2015}, audio models \cite{kell2018task, Freteault2023}, and video-language models (VLMs) \cite{nakagi2024unveiling}, to predict brain activity.

However, existing studies face challenges in generalizability and comparability. Differences in stimulus modality, quantity, and content, as well as in preprocessing and scoring, make cross-study comparisons difficult.

The Algonauts 2025 Challenge \cite{gifford2024algonauts} provides a framework to address these issues, offering an openly available, preprocessed dataset with a large amount of data per subject and aligned stimuli across modalities, including video, audio, and transcripts, along with a standardized evaluation procedure. The challenge places particular emphasis on generalizability, including both in-distribution and out-of-distribution test sets to rigorously evaluate how well models transfer to new stimuli.

In this report, we present our submission to the competition, based on a multi-modal encoder, which combines representations from pre-trained models across modalities to predict brain activity. To do this, we probed representations from text (Llama-v3-8B\cite{meta_llama31_report}, SmolMv2-1.6B \cite{allal2025smollm2}, Qwen-2.5-7B \cite{qwen2025qwen25technicalreport}), video (slow\_r50 \cite{feichtenhofer2019slowfast}, ViViT~\cite{arnab2021vivit}, VideoMAE~\cite{tong2022videomae}), audio (Whisper-small, Whisper-large \cite{radford2022whisper}), and text–vision (InternVL-1B \& 8B \cite{zhu2025internvl3exploringadvancedtraining}) neural networks. We also experimented with enhancing the transcripts to extract richer representations from pretrained neural network, and adapting vision (slow\_r50) and language models (Llama-3.1-7B) through both stimulus- \cite{Freteault2023} and fine-tuning \cite{merlin2024languagemodelsbrainsalign} to improve brain prediction accuracy. Predictions based on individual models are then integrated via stacked regression \cite{lin2024stacked, antonello2023scaling}. 

Our approach, submitted under the team name Seinfeld, achieved 10th place (out of 27 submissions in the second stage). 
In addition to describing our model and its components, we also share insights from preliminary experiments and compare it with top-performing entries. All code has been made publicly available.

\section{The Algonauts 2025 Challenge}

The goal of the challenge was to predict the brain activity of four subjects in response to a wide range of movie stimuli. Their brain activity was recorded as part of the Courtois Project on neural modeling (CNeuromod, \cite{boyle2020courtois}), and all four subjects opted to release their recordings publicly. Models could be fit on the previously released data of the project, which spans ~65h of movie watching for each of the four subjects. This forms the \textbf{training set} of the challenge and includes data for 'Friends' seasons 1-6, three feature length movies ('Hidden Figures', 'Bourne' and 'The Wolf of Wall Street') and one BBC nature documentary ('Life').

The organizers provide movie stimuli, time-aligned transcripts, and preprocessed fMRI data (1000 Schaefer parcels \cite{schaefer2018localglobal} in MNI152 space), along with baseline features (extracted from three neural networks) and a code base to derive the baseline model (for details see \cite{gifford2024algonauts}). Prediction performance is quantified as the (per-parcel) Pearson correlation of real and predicted MRI time-series on held out test sets, averaged across parcels and then subjects. The competition encompassed two separate test sets: 

The \textbf{Model Building stage} (January to July 2025) featured the \uline{in-distribution 
 test set} from 'Friends' season 7. Stimulus data was available throughout this stage, and teams could upload predictions to be scored on the challenge platform (Codabench).

The \textbf{Model Selection stage} (July 6-13) followed immediately after, introduced the 
\uline{out-of- distribution test set}, consisting of clips from previously unknown movies and series: 'The Pawnshop', 'World of Tomorrow', 'Princess Mononoke', 'Planet Earth', and 'Passe-partout'. Teams had one week to generate predictions on this dataset, and the winner was selected based on their performance.

\begin{figure}[ht]
    \centering
    \includegraphics[width=1\linewidth]{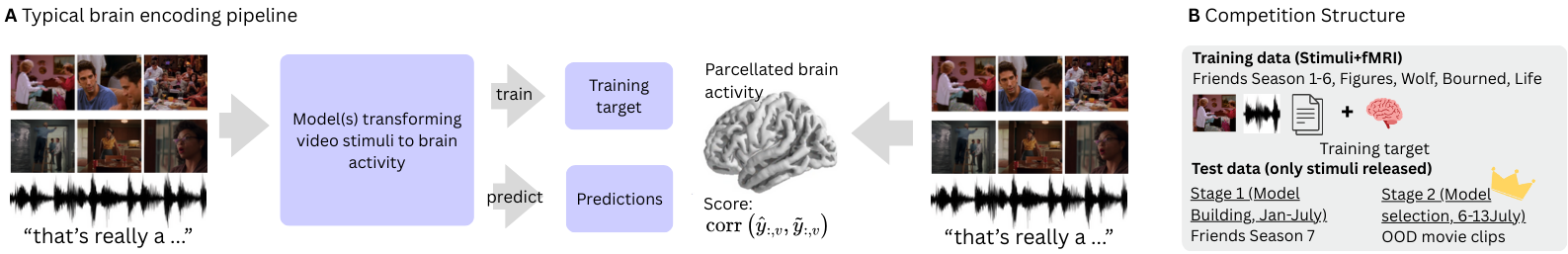}
    \caption{Overview of the brain encoding approach and the competition structure. (A) General fMRI brain activity encoding pipeline. (B) The two main stages of the Algonauts 2025 Challenge, along with the respective training and test datasets.}
    \label{fig:figure1-competition}
\end{figure}

\section{Approach and Results}

We used several strategies to predict brain activity from movie stimuli. The main approaches we explored are:

\begin{enumerate}
  \setlength{\itemsep}{-4pt} 
  \item Using internal representations from pre-trained deep neural networks to fit linear models (Section \S\ref{sec:vanilla})
  \item Enhancing transcripts to extract richer representations from a vision-language model (Section \S\ref{sec:internvl-transcr})
  \item Stimulus-tuning large language models to improve their internal representations (Section \S\ref{sec:llama-st})
  \item Fine-tuning LLMs to predict brain activity directly (Section \S\ref{sec:llama-ft})
  \item Fine-tuning a vision model (slow\_r50) to predict brain activity directly (Section \S\ref{sec:slow_r50_ft})
  \item Modeling poorly predicted brain regions using Hidden Markov Models (Section \S\ref{sec:hmm})
  \item Training a contrastive Video-fMRI model (Section \S\ref{sec:contrastive})
\end{enumerate}

Although not all of these models were included in the final submission, we report them here for completeness and transparency.

\begin{figure}[!b]
    \centering
    \includegraphics[width=1\linewidth]{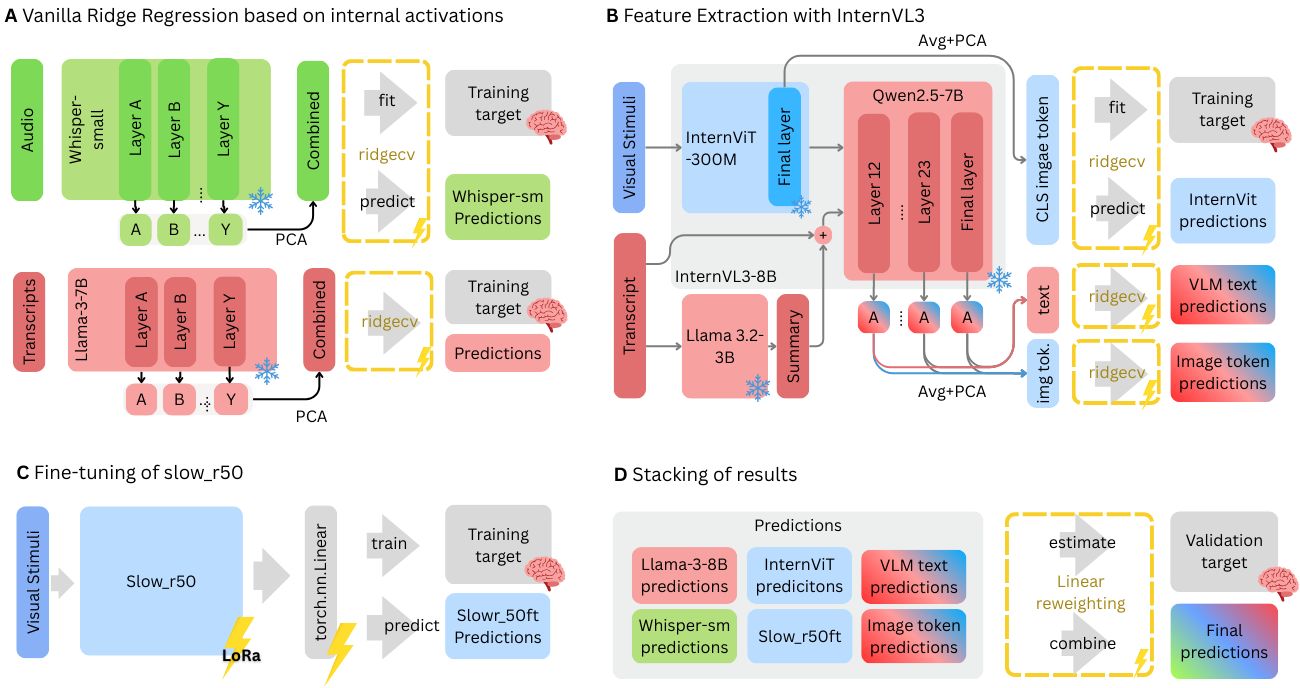}
    \caption{Overview of prediction sources and stacking approach used in our final submission.}
    \label{fig:figure2-complete submission}
\end{figure}

For our final submission, we combined three sources of predictions using stacked regression: linear predictions based on \textbf{representations from two pre-trained deep neural networks} (Llama-3.1-8B, whisper-small), linear predictions based on \textbf{InternVL \cite{zhu2025internvl3exploringadvancedtraining} representations extracted with enhanced transcripts}, and direct \textbf{predictions from slow\_r50 fine-tuned on brain activity}. This combination is shown in Figure \ref{fig:figure2-complete submission}.

In the following sections, we describe each of the tested strategies in detail. The last Section (\S\ref{sec:stacking}) explains how the model predictions were combined for our final submission. We conclude the report with a brief discussion.

\subsection{Using internal representations from pre-trained deep neural networks} \label{sec:vanilla}

The strategy of using internal representations of pretrained neural networks to predict brain activity (\ref{fig:figure2-complete submission}A) has been widely employed in the literature and is also used to generate the Algonauts baseline. In addition to the baseline models, we curated a set of state-of-the-art audio (audio encoder of both whisper-small and whisper-large), vision (slow\_r50) and language deep neural networks (Llama 3.1-8B, SmolLM2-1.7B, Qwen 2.5-7B), and assessed how well their internal representation are predictive of brain activity. For that we first extracted their internal representations to the movie stimuli, whereby the exact extraction process differed between model types. Briefly, for the audio and speech models we used a context length of 10-19 seconds (to mirror training setups for the different models) and collected the representations for all tokens corresponding to the current TR at 4 equally distributed encoder layers. For the vision models, we sample one image per second (or 8 to match the video models) and forward it through the deep neural network and collect the representations of all patches at 4 equally distributed layers. For large language models, we use a context length of 1000 words (ca. 1000-1200 tokens, depending on the tokenizer) and collect representations from the last seven input tokens (double the average number of tokens per TR) at 4 equally distributed layers. 

We flatten (across token/patches and layers) and then concatenate these representations across TRs. We then estimate PCA components based on every fifth sample (for a less computationally intensive estimation) from the held-out Friends season 7 stimuli and then reduce the full feature time series of each neural network model to 2000 dimensions. 

These reduced representations we then used as predictors in the linear encoding pipeline provided by the alognauts competition (sklearn ridge-cv, Leave-one-out cross-validation, with one alpha per output dimension). To identify the best models per modality, we systematically train and test encoding models while varying the number of features/components retained ($n_{comp}$) and the stimulus window ($sw$ in range 1-3). We assume a fixed minimal hemodynamic response delay of 3TRs (ca. 5s), so that for a hypothetical $sw$ of 2, 
we use the concatenated representations for the stimuli presented at the two TRs immediately preceding the 5s hemodynamic delay 
($t-5\,\text{s}$ and $t-6.5\,\text{s}$)
as predictors for brain activity at time t.

For identification of the best model we focus on a single subject (sub-01) and train on Friends seasons 1-5 and test on season 6 ($f6$) and the Hidden Figures movie ($hf$). Based on these experiments, we retained whisper-small and Llama-3.1-8B for the final model. Slow\_r50 showed also high encoding performance, but we ultimately used a superior fine-tuned version described in Section \ref{sec:slow_r50_ft}.

Best parameters and performance scores for the selected models are shown in Table~\ref{tab:vanilla_model_performance}, and the full results across the tested models are shown in Appendix \ref{sec:vanilla_regressions_append}. By visually inspecting the per-parcel correlation performance, we further ensured that they capture activtiy in different cortical regions, indicating that combining them at a later stage might boost performance.

\begin{table}[h!]
\centering
\begin{tabular}{lcccc}
\toprule
Model & $n_{\text{feat}}$ & $sw$ & $r_{f6}$ & $r_{hf}$ \\
\midrule
whisper-small* & 2000 & 3 & \textbf{0.196} & \textbf{0.179} \\
slow\_r50 & 100 & 3 & 0.156 & 0.139  \\
Llama-3.1-8B* & 500 & 3 & 0.162 & 0.148 \\

\cmidrule(lr){1-5} 

InternVL3-8B (Vision Tokens - InternVIT)* & 250 & 2 & 0.164 & - \\
InternVL3-8B (Vision Tokens - Qwen)* & 250 & 2 & 0.174 & - \\
InternVL3-8B (Language Tokens - Qwen)* & 250 & 2 & \textbf{0.189} & - \\

\cmidrule(lr){1-5} 
Llama-3.1-8B & 500 & 3 & 0.162 & \textbf{0.148} \\
Llama-3.1-8B (stimulus-tuned) & 2000 & 3 & \textbf{0.165} & 0.142 \\

\cmidrule(lr){1-5} 
slow\_r50 & 100 & 3 & 0.161 & -  \\
slow\_r50 (fine-tuned)* & 100 & 3 & \textbf{0.178} & -  \\

\bottomrule
\end{tabular}
\captionsetup{width=.7\linewidth}
\caption{Selected models, best parameters and performance scores on held out data for sub-01. Predictions included in the final stacking model are marked with an asterix (*). Highest scores per section are marked in bold.} 

\label{tab:vanilla_model_performance}
\end{table}

\subsection{Enhancing transcripts to extract richer representations from InternVL3}
\label{sec:internvl-transcr}

Given the good results we saw with both vision and language models, we decided to explore if performance can be further enhanced using a multimodal model that combines both modalities. For that we selected the vision-language model (VLM) InternVL3\cite{zhu2025internvl3exploringadvancedtraining} because of its state-of-the-art benchmark results and open source availability. 

For part of our intial experiements, we used InternVL3-1B which uses InternVIT-300M\cite{vit} for vision processing followed by Qwen 2.5 0.5B\cite{qwen2025qwen25technicalreport} as the LLM backbone. However, later experiments and our final submission to Algonauts 2025 used the larger InternVL3-8B which is based on InternVIT-300M and Qwen2.5-7B. We fit linear models for all subjects on 'Friends' seasons 1-5 and tested on season 6, predicting brain activity from the extracted latent representations. This final approach is depicted in Figure \ref{fig:figure2-complete submission}B.

We focused on two key strategies for improving brain prediction accuracy: enriching the transcripts and trying out different layer combinations from which we extract the representations.

First, we demonstrated improvements in encoding accuracy by providing Language Models with more contextual information through enhanced transcripts. Rather than using the simple word transcripts originally provided, we created detailed transcripts for 'Friends' seasons 1-7 that include prepended scene summaries, speaker names, and non-spoken contextual information. An example of the enhancements can be seen in Box \ref{box:suppl_det_vs_simpl}. We hereby use the representation at the final layer of the language backbone (language\_model.model.norm) reduced to 250 dimensions using PCA as basis to predict brain activity. For that, we again use the default Algonatus regression approach with a stimulus window of 2 and hemodynamic response delay of 3TRs, providing only the respective textual (capped to a maximum context of 1000 words) and no vision inputs. 

This results in an average encoding performance (across all 4 subjects) of $r_{f6}$=0.127 for the detailed transcripts without scene summaries, and $r_{f6}$=0.136 with the scene summaries, compared to $r_{f6}$=0.121 when only using the simple transcripts.

To then make use of the integrative properties of the VLM, for the next set of experiments we used concurrent visual and text inputs. As text inputs we used the detailed transcripts, and as vision inputs we used the 8 frames sampled during the TR. Preliminary experiments (see Supplementary Section \ref{sec:suppl_vlm_text_img_token_order}) based on InternVL3-1B showed that interleaving image tokens between transcript segments corresponding to the previous TRs and the current TR had better performance compared to placing image tokens before the transcript segment. In the final model, we input the summary of the transcript for the previous TRs first, followed by the image tokens, and the transcript for the current TR.

We next sought to determine which layers embeddings were most predictive of the brain activity. For that we probed various layers of both the image and language backbone of InternVL3-8B.
The full set of experiments using different token and layer combinations are described in Supplementary Table \ref{tab:vlm_final_embd_perf}. Typically, we first extracted the representations for different tokens and layers for each TR, in some cases averaged across tokens, and then concatenated them. Afterwards, we reduced the concatenated representations to 250 dimensions using PCA and finally fit a linear model as previously described.

We found that using the representations averaged across the last 10 text tokens at layers 12, 21, 22 and 23 of the language backbone, \emph{Language Tokens (Qwen)}, performed best ($r_{f6}$=0.195 across subjects). This is shortly followed by using the average image tokens (averaged across all tokens for each of the 8 input image frames) at the norm layer of the language backbone, \emph{Vision Tokens (Qwen)}, with $r_{f6}$=0.178 across subjects, and the CLS tokens for each of the 8 input image frames at the norm layer of the vision backbone, \emph{Vision Tokens (InternViT)}, with $r_{f6}$=0.168 across subjects. For the full layer names see Box \ref{box:suppl_vlm_emb_layers}. The results of these key combinations for \emph{only sub-01} are also included in Table \ref{tab:vanilla_model_performance} for comparison. In our final submission we include both the predictions based on \emph{Language Tokens (Qwen)} and \emph{Vision Tokens (Qwen)}.

\subsection{Stimulus-tuning Llama-3-7b on Friends transcripts}
\label{sec:llama-st}

We hypothesized that adapting Llama-3-7b specifically to movie dialog structure might further improve brain prediction accuracy. Our reasoning was that if the model is trained on predicting data that more closely resembles movie dialog structure, it might better capture relevant representational dimensions.

For this, we used the 4-bit quantized version of Llama-3-7b-base and employed continuous pretraining to allow for flexible adaptation of the LLM outputs to the specific dialog structure from our dataset.

As the basis, we used the per-TR transcripts\footnote{TR (Time of Repetition): brain imaging acquisition timepoints. In the Algonauts dataset, the scanning sequence was configured such that acquiring one complete brain volume (=brain image) took 1.49s. Per-TR transcripts contain words spoken during each TR interval.} provided by the Algonauts competition. Prior to stimulus-tuning, we concatenated words across TRs, retaining the most recent 500 words leading up to the current TR to balance sufficient context with computational constraints during fine-tuning. To approximate division into scenes and dialog turns that were absent from the provided transcripts, we replaced 1-2 empty TRs with a single line break and longer silences ($>2$ TRs) with two line breaks. An example for this can be seen in Box \ref{box:suppl_transcr_stim_tune} 

Using the thus prepared transcripts, we then continued pretraining for 1 epoch based on 'Friends' S1-5. For this, we employed Low Rank Adaptation (LoRA)\cite{hu2021lora} with a lora rank of 128, a lora alpha of 32, a batch size=2, a gradient accumulation step size of 8, and a learning rate of 5e-5 with the AdamW8bit optimizer using a cosine learning rate schedule. This results in 4.19\% of parameters being trained, which could be conducted using a single google colab GPU. This stimulus tuning resulted in dialogues that indeed seemed to approximate the structure and contents of 'Friends' dialogues (shown in Box~\ref{box:suppl_base_x_stim_tuned_llm}).

After training was completed, we followed the same procedure as in Section \S\ref{sec:vanilla} for extraction of the internal representation, dimensionality reduction, and regression of brain activity. The stimulus-tuned model showed slightly elevated scores (again after systematic variation of $n\_{feat}$ and $sw$) for 'Friends' season 6 ($r_{f6}=.165$ vs $r_{f6}=.162$ before stimulus-tuning) but reduced scores for 'Hidden Figures' ($r_{hf}=.142$ vs $r_{hf}=.148$ before stimulus-tuning). This shows that prediction performance may benefit from stimulus-tuning, but does not necessarily generalize to out of distribution test sets.  Given the mixed and modest nature of these performance changes, we did not include this approach in our final submission. 

\subsection{Fine-tuning of Llama-3.2-3B for fMRI prediction}
\label{sec:llama-ft}

Next, we wanted to see whether directly fine-tuning the LLM to predict the fMRI data can boost performance. We again used LoRA adapters on each layer (lora rank=128, lora\_alpha = 32, batch\_size=32, grad\_accum=4, using AdamW8bit and a cosine learning rate schedule), adapting the following layer components: 'q\_proj', 'k\_proj', 'v\_proj', 'o\_proj', 'gate\_proj', 'up\_proj' and 'down\_proj'. We added a single linear layer that predicted activity at each of the 1000 parcels, taking as input the concatenated hidden states from 4 transformer layers, which were equally spaced throughout the backbone to capture a maximum of variance.

We collected the activations for the 10 last tokens, giving the preceding 100 words as context to the LLM (the shorter context length enabled larger batch size and faster training).  To account for the hemodynamic response delay, we shifted the text stimuli by 4 TRs (6s).  We train on 'Friends' season 1-4 and 6, 'Hidden Figures' and test on every 10th TR on the test set ('Friends' season 5, 'The Wolf of Wall Street' and 'Life'). 
We found that testing on every 10th TR approximates the score of testing on the full timeseries, but is substantially faster.\footnote{This was to reduce computational costs. Preliminary experiments comparing different sampling rates (full timeseries, or every 2nd, 5th, 10th, and 100th TR) showed that scores based on every 10th TR correlated at r=0.96 with full timeseries scores, indicating this sampling rate maintains evaluation accuracy while substantially reducing compute requirements.}

Both LoRA adapters and linear layer were jointly trained, each with separate learning rates. We tried \{4e-06, 8e-06, 2e-05, 4e-05\} as learning rate for the LoRA adapters and \{1,2,4,8\}e05 for the linear layer. The combination of $lr_{lora}=2e^{-05}$ and $lr_{lin}=4e^{-05}$ yielded the lowest validation loss, translating into a score of $r$=0.175 on the combined test set for sub-01. Systematic comparisons across subjects and test sets, and variations to e.g. the layers, shifting, and training parameters, and proper baseline comparisons could not be completed before the deadline. These predictions were therefore not be included into our submission.

\subsection{Fine-tuning a vision model (slow\_r50) to predict brain activity directly}
\label{sec:slow_r50_ft}

Similarly, we sought to increase prediction accuracy by fine-tuning our vision model. The version of slow\_r50 \cite{feichtenhofer2019slowfast} that we used was pretrained on the kinetics-400 dataset, which contains videos depicting 400 human action classes. The fine-tuning setup is depicted in Figure \ref{fig:figure2-complete submission}C.

Given early layers of Convolutional Neural Networks like slow\_r50 are thought to extract low level features while later layers capture more high level abstract features \cite{yosinski2014transferable}, we decided to add LoRA adapters to the last 2 blocks of slow\_r50 (lora rank=8, lora alpha=16, batch size=32). For a full list of layers adapted, refer to the 'Less extensive layer list' in Box \ref{box:suppl_lora-layers}. We furthermore added a single linear layer that predicted activity at each of the 1000 parcels, taking as input the activations from the 'block.5.pool' layer of slow\_r50. Both LoRA adapters and linear layer were jointly trained, each with separate learning rates. In our experiments cosine annealing over 10 epochs from $10^{-4}$ to $10^{-6}$ for the linear layer and $10^{-4}$ to $10^{-7}$ for LoRA worked best.

We trained on \textit{Friends} season 1-5 and tested on season 6 for sub-01. Table \ref{tab:vanilla_model_performance} shows the pearson r-score for sub-01 after fine-tuning for 3 epochs ($r_{f6}$=0.178) compared to baseline (non fine-tuned slow\_r50, with $r_{f6}$=0.161) as provided by the Algonauts challenge. Appendix \ref{sec:slowr50_details} contains a more detailed description of fine-tuning experiments and results.

We noticed overfitting when fine-tuning for more than 3 epochs. In an attempt to reduce the encoutered overfitting, we fine-tuned a single model for all 4 subjects to reduce overfitting but it's performance did not match individual subject fine-tuning. For details on implementation and experiment results, please refer to section \ref{sec:slowr50_details}.

\subsection{Modeling poorly predicted brain regions using Hidden Markov Models}
\label{sec:hmm}

In this experiment, we aimed at leveraging information from the BOLD signal itself, independent of stimulus presentation by 1) predicting the BOLD signal in the test set from its own history in the training set, and 2) using information from the generally better-predicted parcels to improve accuracy in the generally worse-predicted parcels. 

To do this, we employed a Hidden Markov Model (HMM) with two different observation models: 1) a standard Gaussian and 2) a Gaussian, regression-based observation model, the Gaussian-Linear Hidden Markov Model (GLHMM, \cite{vidaurre2025gaussian}). The full algorithm is given in the Appendix \ref{sec:appendix_hmm}. The main difference between these is that the standard Gaussian HMM models the fMRI timeseries as a multivariate Gaussian with a time-varying mean (amplitude) and covariance (functional connectivity), while the GLHMM additionally assumes that the amplitude in some brain areas depends on the amplitude in other areas. The standard Gaussian HMM assumes that an observed timeseries $Y$ was generated by a sequence of K hidden states, each described by a Gaussian distribution with mean $\mu$ and covariance $\Sigma$, i.e. $Y_t \sim \mathcal{N}(\mu^k, \Sigma^k)$ at time point $t$ when state $k$ is active. 

The GLHMM uses the same general observation model, but additionally models the timeseries’ dependency on a second set of observations $X$ as linear regression, so that $Y_t \sim \mathcal{N}(\mu^k + X_t \beta^k, \Sigma^k)$, where $\beta$ is the state-dependent regression parameter. In addition to these state parameters, we model the temporal evolution of states using an initial state probability $\pi$ and transition probabilities $A$. All parameters are estimated using variational inference.

Variational inference is an iterative, two-step procedure that alternates between updating the variational distributions for the hidden states and the model parameters. Step 1: Given the current observation model parameters, we compute the posterior probabilities of the hidden states and transitions using a modified Forward-Backward algorithm, based on the probabilities of the past (forward) and future observations (backward) at each timestep. Step 2: The parameters of the observational model are updated based on those posterior probabilities. For example the means and covariances of the variational distributions are updated based on a weighted average of the data, using the posterior probabilities of each time point's state as the weights. Similarly, if applicable, per state regression betas are estimated using weighted regression. This aims at maximizing the fit of our approximate model to the data, i.e., minimizing the variational free energy. Step 1 and 2 are repeated until convergence.

The timeseries were standardised separately for each scanning session. For the GLHMM, we separated the parcels into a minority of better-predicted parcels (predictor parcels, $X$), here all 80 parcels belonging to the bilateral visual cortices, and a majority of worse-predicted parcels (dependent parcels, $Y$), all remaining parcels. To avoid overfitting, we used principal component analysis to reduce dimensionality of the predictor parcels to 10 principal components (PCs) and the dependent parcels to 100 PCs. For sampling, PCs were backtransformed into original dimensionality. For all models, we use $K=10$ states. To generate predicted timecourses, we sampled from the trained models independent of stimuli. For the GLHMM, we used two sampling/prediction strategies: first, we sampled from the model using the true predictor timecourses (i.e. the measures fMRI timeseries) as $X$, then, to gauge prediction accuracy in a fully held-out test set, we used timecourses predicted (i.e. the predicted timeseries by an auxiliary model; this mirrors better the competition condition, as the ground truth fMRI data is not available for the test set) by the combined out-of-distribution model (see Stacking of model output) (provider model), as predictor timecourses $X$.

As a preliminary exploration, we trained both HMMs on subject 1, season 5 of Friends and tested on season 6. The standard Gaussian HMM achieved a mean accuracy of 0.001, showing little promise for further exploration. The GLHMM based on the true time-courses of the predictor (better-predicted) parcels achieved a mean accuracy of 0.180 on the dependent (worse-predicted) parcels. This suggests it captured some relevant information. We then trained the GLHMM on all Friends seasons and tested on the movie Hidden Figures, achieving an average accuracy of 0.116.

However, when using the predicted time courses of the predictor parcels rather than the true time courses, accuracy dropped to 0.001. This indicates that the approach would only be fruitful if predictions for the predictor parcels from the provider model were extremely accurate, and was therefore not used in the final model.

\subsection{Contrastive Video-fMRI Encoder}
\label{sec:contrastive}

In parallel, we experimented with a different approach to extract features from visual stimuli, training a contrastive Video-fMRI encoder. The intuition is that this approach should guide the model in extracting features from the visual stimuli that correlate more with the fMRI activations. For this, we aligned short video chunks from the Friends dataset (configurable length, defaulting to 32 frames per chunk) to fMRI samples with explicit control over stimulus and fMRI windows and an adjustable HRF  (similarly to \S\ref{sec:vanilla}); temporal (e.g., 2x) and spatial downsampling were also used to reduce memory of video stimuli (e.g., videos were downsampled to 224x224 resolution). For visual inputs, we experimented with two video architectures based on ViViT~\cite{arnab2021vivit} and VideoMAE~\cite{tong2022videomae}, pretrained on Kinetics-400. For fMRI samples, we explored two encoder architectures based on an MLP and on 1D convolutions. The contrastive objective follows a CLIP-style temperature-scaled loss (InfoNCE) to align video and fMRI embeddings in a shared space; regression models predict voxel responses directly. We trained with AdamW (and alternatives) using configurable learning-rate schedules (step or cosine) and optional warmup. Automatic mixed precision (torch.amp) was used to reduce memory and increase throughput. Different configurations of batch size (128, 256, 512), learning rate (initial value in the range [1e-4, 1e-5]), and time sampling (1x, 2x) were explored. Additionally, a windowing approach was tested in which we included previous chunks of stimuli and fMRI activations in the model input (1 to 10 previous chunks were included). Once trained, the encoder was frozen, and a small MLP composed of two linear layers, with batch normalization and dropout ($p=0.3$), was employed to predict the fMRI activations. Evaluation reported per-subject Pearson correlation between model outputs and measured fMRI as the primary metric, with values around $0.1$-$0.15$. We experimented with training a separate model on each subject, or encoding the subject identity as input with a linear embedding layer. The two approaches did not differ significantly in terms of results. Overall, both VideoMAE and ViViT reached similar results, with the MLP encoder for fMRI data performing slightly better than the 1D-convolutional one. Although we believed the approach to be promising, we abandoned it in favour of the previous ones, due to time constraints, as they seemed to perform better.

\subsection{Stacking of Model Outputs}
\label{sec:stacking}
To combine the predictions from our different strategies we used stacked regression \cite{lin2024stacked}, which allows to combine predictions from multiple models by learning optimal linear weights to maximize predictive performance.

We tested various prediction set combinations. For each prediction set combination, we systematically probed different dataset splits for fitting versus stacking optimization.

Our best performing combination was the following: (1) predictions based on whisper-small, (2) predictions based on Llama-3.1-8B, (3) predictions based on vision token representations from the InternVL3 language backbone (\emph{Vision Tokens - Qwen}), (4)  predictions based on language token representations from the InternVL3 language backbone  (\emph{Language Tokens - Qwen}), (5) predictions based on vision token representations from the InternVL3 vision backbone (\emph{Tokens - InternVIT}) and (6) predictions from the fine-tuned slow\_r50 model. This is also shown in Figure \ref{fig:figure2-complete submission}D. 

Testing revealed the following best performing dataset split for this combination: Diverging from our previous experiments (see section \ref{sec:vanilla}), whisper-small and Llama-3.1-8B based linear regression models were only trained on 'Friends' seasons \textbf{1-4} and \textbf{6}. We intentionally left out season 5 to enable less biased estimation of the stacking parameters. We then optimized the prediction stacking parameters based on 'Friends' season \textbf{5}, '\textbf{B}ourne' and '\textbf{W}olf of Wall Street'. This dataset split can be summarized by the shorthand \textbf{"12346-5BW"} - using season numbers and first letters of movie titles to signify datasets, with the dash separating fitting from stacking datasets. This achieved a performance of 0.274 for sub-01 compared to the best dataset split ("12345BW-BW") for the combination of the linear predictions based on representations from the three unaltered pre-trained neural networks (whisper-small, Llama-3.1-8B, and slow\_r50) alone, which achieved an r=0.221 in the same test.

For our final submission, we included also 'Hidden Figures' for model fitting ("12346\textbf{F}-5BW"), as we reasoned that incorporating more non-'Friends' data into the fitting set would enable a better generalization. This led to our final score of 0.1496 on the competition leaderboard for the out-of-distribution test set averaged across subjects. 

\section{Discussion}

\textbf{Fine-tuning of deep neural networks can help, but training a separate backbone that takes pre-extracted internal representations as input is more efficient.} We demonstrated performance boosts through fine-tuning slow\_r50 and attempted similar improvements with stimulus and fine-tuning Llama (with mixed results). However, the top 3 teams \cite{d2025tribe, Team2_NCG, eren2025multimodal}, achieved greater efficiency by training separate non-linear architectures - either transformers \cite{d2025tribe,Team2_NCG} or 2-stage RNNs \cite{eren2025multimodal} - that take multiple pre-extracted embeddings as input. This approach led to substantial performance boosts, with the winning team improving from 0.23 to 0.29 on Friends season 7. This demonstrates that the pre-extracted hidden states already contain much more information relevant for brain encoding than may be revealed by linear models, and need not necessarily be fine-tuned first. The efficiency of training on pre-extracted features (i.e. much less architecture needs to be passed through at each forward pass and likely fewer trainable parameters) allows these teams to run hundreds or thousands of training runs with different initializations and ensemble the results, further boosting performance.

\textbf{Overfitting is a common issue but may be remedied.} We initially also experimented with a transformer encoder (8 attention heads, 2 layers) using pre-extracted features from the Algonauts challenge (BERT, audio via MCC, and slow\_r50 video embeddings), which resulted in overfitting and poor performance. However, successful teams overcame similar challenges, likely through architectural design choices. TRIBE \cite{d2025tribe} used larger dimensions per modality and added learnable cross-subject layers after their transformer architecture. 
The third-placed team \cite{eren2025multimodal} also reported overfitting challenges but employed a 2-stage RNN approach with positional embeddings which may have been key in resolving the issue.  

\textbf{Additional performance gains can be realized through cross-subject models, but implementation matters.} We experimented with training a single cross-subject model when fine-tuning slow\_r50. However, our cross-subject model could not match the performance of per-subject fine-tuning. In contrast, TRIBE adds a cross-subject learnable layer after their cross-modal transformer and experienced a performance boost. 

\textbf{Neural networks of different modalities encode complementary brain-encoding-relevant information, but adding multiple neural networks of the same modality only leads to marginal improvements.} Predictions based on deep neural networks encode complementary information that is predictive of different brain areas, though they also show partial overlap across modalities (e.g., Llama best encoded auditory and language areas, but was also somewhat predictive of the Visual Network despite not receiving any visual input). Adding further neural networks from the same modalities only led to marginal improvements. We found the same when experimenting with stacking different prediction sets. For example, even though slow\_r50 (Convolutional Neural Network based) and InterVIT (Transformer based) have different base architectures, stacking them together or substituting one for another did not alter prediction performance markedly. This aligns with observations from \cite{Team2_NCG}. 

\textbf{OLS/Ridge based linear models may underestimate brain encoding capacity of deep neural networks.} As shown by the fourth-placed team \cite{villanueva2025predicting}, optimizing more complex linear models using AdamW and backpropagation can far exceed PCA+OLS approaches (including all of our experiments based on linear approaches) and even become competitive with top non-linear encoding models \cite{scotti2025insights}. Nonetheless, Ridge-based approaches remain widely used in the field. Different regularization strategies have also been suggested \cite{nunez2019banded} to improve prediction performance, though they can be computationally intensive.

\textbf{Brain encoding performance of deep neural networks can be improved by augmenting/preprocessing the stimuli.} Another notable contribution of our work was showing that modifying the textual inputs increases VLM brain prediction performance. Adding detailed context further enhances the extracted representations, consistent with \cite{nakagi2024unveiling}. Future work could explore how different networks are affected by varying the type and amount of contextual detail.

\textbf{Open questions.} Several directions remain unexplored. First, we did not test whether different approaches such as fine-tuning and enhancement of the transcripts may complement each other synergistically. Second, it's unclear whether linear and non-linear methods capture distinct information or merely different views of the same underlying structure. Further, our fine-tuning experiments with LLaMA and slow\_r50 suggest that non-linear feature combination approaches (as used by other teams) might also benefit from end-to-end training of both the backbone and the combination layers, rather than extracting fixed features from frozen backbones. However, the computational costs must be weighed carefully. More broadly, identifying architectural and training designs that yield richer, more brain-like representations remains a challenge for future work.

\section{Code Availability}
\label{sec:code_sharing}
For sections 2.1 and 2.2.1, 2.2.2 and 2.4 refer to the following repo:\\
\href{https://github.com/rscgh/algo25}{https://github.com/rscgh/algo25}

For fine tuning slow\_r50 and InternVL 3 feature extraction please refer to:\\
\href{https://github.com/bagga005/algonauts}{https://github.com/bagga005/algonauts}

For the detailed transcripts that were used for InternVL 3, please refer to:\\
\href{https://github.com/bagga005/friends_transcripts_algonauts25}{https://github.com/bagga005/friends\_transcripts\_algonauts25}\\


\bibliography{references}     

\newpage
\appendix

\renewcommand{\thetable}{S\arabic{table}} 
\setcounter{table}{0}

\renewcommand{\thefigure}{S\arabic{figure}} 
\setcounter{figure}{0}                       

\setcounter{roundedboxctr}{0} 
\renewcommand{\theroundedboxctr}{S\arabic{roundedboxctr}}


\hrule height 1pt
\section*{Supplementary Materials}
\vspace{1mm}
\hrule height 1pt
\vspace{4mm}


\section{Vanilla regression experiments}

\subsection{Model performance}

\label{sec:vanilla_regressions_append}

\begin{figure}[htp]
    \centering
    \includegraphics[width=1\linewidth]{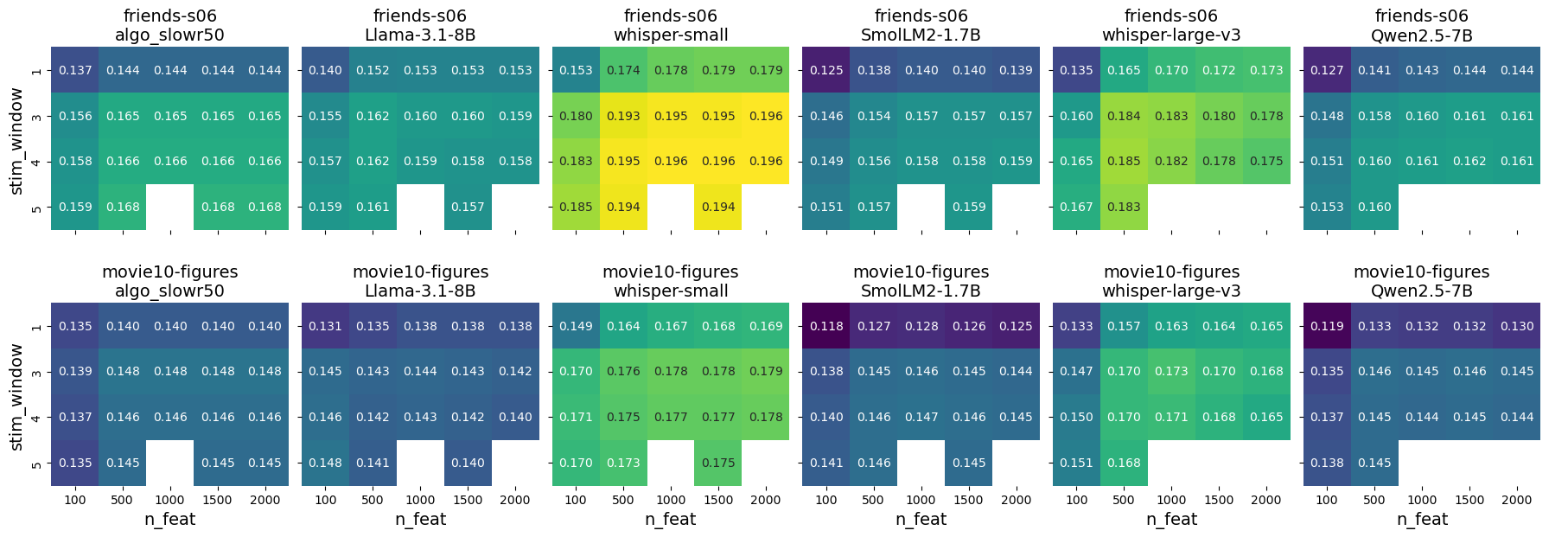}
    \caption{Enter Caption}
    \label{fig:placeholder}
\end{figure}



\subsection{Transcripts used for stimulus and fine-tuning of language models}
\label{appendix:stim_tuning-transcripts}

\begin{center}
\refstepcounter{roundedboxctr} 
\label{box:suppl_transcr_stim_tune} 
\begin{roundedbox}[halign=flush left]{Transformed transcripts for stimulus tuning}
\noindent\textbf{Before:} Hey, Rachel. Hi. What a nice surprise. What are you doing here? Well, you know, I was just in the neighborhood and I passed by your building and I thought to myself, what's up with Carol and sweet little Ben? Nice. Come on in. Okay. I'll make some coffee and we can chat. I'd love that. I would love that. So where is sweet little Ben? I would love to have a little. I found him. Very funny. Come here. That is exactly why I've come here to talk to you. Okay. Rach, do you want some sugar in your coffee? Yes, but do I want sugar in my coffee? 
\tcblower
\noindent\textbf{Transformed:} Hey, Rachel. Hi. What a nice surprise. What are you doing here? Well, you know, I was just in the neighborhood and I passed by your building and I thought to myself, what's up with Carol and sweet little Ben?\textcolor{red}{\_}  

Nice. Come on in. Okay. I'll make some coffee and we can chat. I'd love that. I would love that.\textcolor{red}{\_}  

So where is sweet little Ben? I would love to have a little.\textcolor{red}{\_}

I found him. Very funny. Come here. That is exactly why I've come here to talk to you. Okay. Rach, do you want some sugar in your coffee? Yes, but do I want sugar in my coffee?\textcolor{red}{\_} 
\end{roundedbox}
\end{center}


\section{InternVL Implementation Details and Results}
\label{sec:vlm_details}
\subsection{Detailed Transcripts}
A simple (list of words spoken) transcript was provided in the SDK as part of the Algonauts 2025 challenge. In order to provide better context to the LLM layer, we prepared detailed transcripts for \textit{Friends} seasons 1-7. In addition to words spoken, detailed transcript contains 1) a brief description of every scene 2) speaker name 3) context such as [stares in disbelief] in Box ~\ref{box:suppl_det_vs_simpl}. 

\begin{center}
\refstepcounter{roundedboxctr} 
\label{box:suppl_det_vs_simpl} 
\begin{roundedbox}[halign=flush left]{Simple and Detailed Transcripts}
\noindent\textbf{Simple Transcript} \\
hey eddie you uh wanna play some foosball no thanks man I'm not uh I'm not really into sports yeah ok alright doesn't matter time for baywatch
\tcblower
\noindent\textbf{Detailed Transcript} \\
{[}Scene: Chandler and Eddie's apartment.{]} \\
CHANDLER: Hey Eddie, you uh, wanna play some foosball? \\ 
EDDIE: No thanks man, I'm not uh, I'm not really into sports. \\
CHANDLER: [stares in disbelief] Yeah o-, OK, alright. [oven timer goes off] Doesn't matter, time for Baywatch.
\end{roundedbox}
\end{center}

We obtained detailed transcripts from \cite{friends_det_trans} and then did fuzzy matching to match the detailed transcript with the simple transcript provided in SDK. Table~\ref{tab:suppl_dettrans_perf} shows performance comparison of SDK provided simple transcript as input to Llama 8B and detailed transcript as input to Qwen 2.5 1B. In previous tests we had already seen Llama 8B outperform Qwen 2.5 1B on equivalent data. In our experiment, we also tried a blended transcript - a mix of simple (43\%) and detailed transcript (57\%). The blended transcript would use detailed transcript if our fuzzy matching found a confident match, else use the simple transcript. Given this experiment was to measure relative performance gain of detailed transcripts, we only used the last token from language norm layer of Qwen.

In order to provide concise context to LLMs, we prepared summaries for each episode. Multiple summaries were prepared for each episode. A summary was prepared from start of episode till start of each scene (after the first 1000 words). We used Llama 3.1 8B \cite{meta_llama31_report} to prepare summaries. Table~\ref{stab:vlmsum_perf} shows performance gains made when summaries are prepended to words spoken in a tr. 

We have shared our Detailed Transcripts, Summaries and source for preparing them. Please refer to section \ref{sec:code_sharing} for details.

\begin{center}
\begin{table}[ht]
\centering
\begin{tabular}{lccc}
\toprule
\textbf{Network} & \textbf{Simple Transcript} & \textbf{Blended Transcript} & \textbf{Detailed Transcript} \\
\midrule
Full Brain     & 0.121 & 0.127 & 0.137 \\
Visual Only    & 0.141 & 0.146 & 0.168 \\
\bottomrule
\end{tabular}
\caption{Pearson correlations for \textit{Friends} Season 6 when extracting features based on Simple Transcript, Blend of Simple and Detailed and Detailed Transcript}
\label{tab:suppl_dettrans_perf}
\end{table}
\end{center}

\begin{table}[h!]
\centering
\begin{tabular}{lccc}
\toprule
\textbf{Network} & \textbf{Simple Transcript} & \textbf{Detailed Transcript} & \textbf{with Summaries}\\
\midrule
Full Brain     & 0.121 & 0.137 & 0.139 \\
Visual Network     & 0.141 & 0.168 & 0.171 \\
Somatomotor Network     & 0.099 & 0.115 & 0.119 \\
Dorsal Attention Network    & 0.136 & 0.154 & 0.155 \\
Ventral Attention Network     & 0.107 & 0.125 & 0.128 \\
Limbic Network    & 0.046 & 0.05 & 0.048 \\
Control Network     & 0.116 & 0.121 & 0.12 \\
Default Mode Network    & 0.148 & 0.167 & 0.167 \\
\bottomrule
\end{tabular}
\caption{Pearson correlations for Friends Season 6 for subject 1 when extracting features based on Simple Transcript, Detailed Transcript, Detailed Transcript with Summaries using InternVL 3 1B}
\label{stab:vlmsum_perf}
\end{table}

\subsection{Text and image token input order for the VLM}
\label{sec:suppl_vlm_text_img_token_order}
InternVL adds output from its vision layer as tokens to the LLM and processes them along with the text input. Our vision input to the VLM were 8 equi-spaced images from a given TR (as per Algonauts SDK) at a resolution of 448px by 448px. 

Our inputs to the LLM layer were in the format: \textit{Pre-Text} - Vision Tokens  - \textit{Post-Text}. Here \textit{Pre-Text} refers to n words that are part of transcript prior to current tr. \textit{Post-Text} refers to p words spoken in current tr. And n + p =1000. We used 1000 as it worked well in our LLM experiments. Vision Tokens are output from the vision layer. Box ~\ref{box:suppl_vlm_text_img_token_order} shows sample input to the LLM.

\begin{center}
\refstepcounter{roundedboxctr} 
\label{box:suppl_vlm_text_img_token_order} 
\begin{roundedbox}[halign=flush left]{Input format to InternVL}
\noindent\textbf{Pre-Text} \\
\textbar Scene: Monica and Phoebe’s, three years earlier, Phoebe, Monica, and Ross are there \textbar \\
Phoebe: Oh, really? \\
Ross: Yeah, y'know how I have you guys, well she doesn’t really have any close friends that are just hers, but last week she meet this woman at the gym, Susan something, and they really hit it off, and I-I-I think it’s gonna make a difference \\
\textbar Present Scene: Chandler’s, Chandler is interviewing a potential roommate. \textbar \\
Chandler: Soo, ah, Eric, what kind of photography do ya do? \\
Eric: Oh, mostly fashion, so …
\tcbline
\noindent\textbf{Vision Tokens} \\
Frame1: $<\text{image}>$ \\
Frame2: $<\text{image}>$ \\
Frame3: $<\text{image}>$ \\
Frame4: $<\text{image}>$ \\
Frame5: $<\text{image}>$ \\
Frame6: $<\text{image}>$ \\
Frame7: $<\text{image}>$ \\
Frame8: $<\text{image}>$ \\
\tcbline
\noindent\textbf{Post-Text} \\
Eric:  … there may be models here ...
\end{roundedbox}
\end{center}

\subsection{Selection of VLM layer and token combinations}
We experimented with obtaining activations from different Vision and LLM latent layers. Our thoughts were to get embeddings from early, middle and final layers to get information at different abstraction levels. We obtained embeddings from layers listed in Box ~\ref{box:suppl_vlm_emb_layers}

\begin{center}
\refstepcounter{roundedboxctr} 
\label{box:suppl_vlm_emb_layers} 
\begin{roundedbox}[halign=flush left]{Layers of InternVL3-8B considered for embedding extraction (along with token modality)}
\textbf{Vision Tokens from the vision backbone (based on InternVIT-300M):} 
vision\_model.encoder.layers.\{2,4,12,22,23,norm\} \\

\textbf{Language and vision tokens from the language backbone (based on Qwen2.5-7B):} 
language\_model.model.layers.\{4,12,20,21,22,23,norm\}
\end{roundedbox}
\end{center}

For Vision layers we extracted CLS for the 8 images from the different layers. For the LLM layers we extracted tokens for - Last n \textit{Pre-Text} tokens, Average of all \textit{Pre-Text} tokens, Average of each Image Frame, Token right after the last image frame, Last n \textit{Post-Text} tokens, Average of last n \textit{Post-Text} tokens. Table \ref{tab:vlm_final_embd_perf} shows prediction performance on Friends season 6 after training on seasons 1-5 across subjects using InternVL 3 8B. For each experiment embeddings from different layers are concatenated and PCA is done to select top 250 features.


\begin{longtable}{p{1cm}p{1cm}p{1cm}p{1cm}p{1cm}>{\raggedright\arraybackslash}p{8cm}}
\caption{Pearson correlations for Friends Season 6 for all subjects using embeddings from Vision and Language tokens from Qwen and InternVIT when using InternVL 3 8B \newline Embeddings marked with * are used in competition submission} \label{tab:vlm_final_embd_perf}
\\

\toprule
\textbf{Average} & \textbf{Subj 1} & \textbf{Subj 2} & \textbf{Subj 3} & \textbf{Subj 5} & \textbf{Embeddings} \\
\midrule
\endfirsthead

\toprule
\textbf{Average} & \textbf{Subj 1} & \textbf{Subj 2} & \textbf{Subj 3} & \textbf{Subj 5} & \textbf{Embeddings} \\
\midrule
\endhead

\bottomrule
\endfoot
0.219 & 0.211 & 0.222 & 0.242 & 0.201 & 
Language Tokens (Qwen): Average of up to last 10 Post Text tokens taken for layers norm, 23, 22, 21, 12. \newline Vision Tokens (Qwen): 1 token for each of 8 images taken by taking average of all tokens for that image from layer norm.  \\
0.195 &	0.189 &	0.203 &	0.211 &	0.176 &	
\textbf{*Language Tokens (Qwen): Average of up to last 10 tokens that are part of words in TR taken for layers norm, 23, 22, 21, 12} \\
0.192 &	0.186 &	0.2 &	0.208 &	0.173 &	
Language Tokens (Qwen): Average of up to last 10 tokens that are part of words in TR taken for layers norm, 23, 22, 21 \\
0.186 &	0.179 &	0.192 &	0.204 &	0.17 &	
Language Tokens (Qwen): Average of up to last 10 tokens that are part of words in TR taken for layers 12 \\
0.182 &	0.176 &	0.19 &	0.197 &	0.165 &	
Language Tokens (Qwen): Average of up to last 10 tokens that are part of words in TR taken for layer norm \\
0.181 &	0.174 &	0.189 &	0.195 &	0.164 &	
Language Tokens (Qwen): Average of up to last 7 tokens that are part of words in TR taken for layer norm \\
0.178 &	0.174 &	0.174 &	0.2 &	0.165 &	
\textbf{Vision Tokens (Qwen): 1 token for each of 8 images taken by taking average of all tokens for that image from layer norm} \\
0.177 &	0.171 &	0.184 &	0.193 &	0.161 &	
Language Tokens (Qwen): Average of all tokens that are part of words in TR taken for layer norm \\
0.176 &	0.169 &	0.184 &	0.191 &	0.159 &	
Language Tokens (Qwen): Average of up to last 3 tokens that are part of words in TR taken for layer norm \\
0.174 &	0.167 &	0.181 &	0.189 &	0.158 &	
Language Tokens (Qwen): Average of last token that is part of words in TR taken for layer norm \\
0.169 &	0.161 &	0.176 &	0.185 &	0.155 &	
Language Tokens (Qwen): Average of up to last 10 tokens that are part of words in TR taken for layer 4 \\
0.168 &	0.163 &	0.163 &	0.19 &	0.155 &	
Vision Tokens (InternVIT): CLS tokens for each of the 8 images from layers 12 and 4 \\
0.168 &	0.164 &	0.163 &	0.19 &	0.154 &	
Vision Tokens (InternVIT): CLS tokens for each of the 8 images from from layer 23 \\
0.168 &	0.164 &	0.163 &	0.19 &	0.154 &	
\textbf{*Vision Tokens (InternVIT): CLS tokens for each of the 8 images from layer norm} \\
0.167 &	0.163 &	0.162 &	0.19 &	0.154 &	
Vision Tokens (InternVIT): CLS tokens for each of the 8 images from layer 22 \\
0.161 &	0.156 &	0.158 &	0.179 &	0.149 &	
Language Tokens (Qwen): 1 token that is right after the last image token from layer norm \\
0.147 &	0.141 &	0.144 &	0.167 &	0.137 &	
Vision Tokens (InternVIT): CLS tokens for each of the 8 images from layer 12 \\
0.117 &	0.113 &	0.115 &	0.131 &	0.11 &	
Vision Tokens (InternVIT): CLS tokens for each of the 8 images from layer 4 \\
0.101 &	0.095 &	0.098 &	0.114 &	0.096 &	
Vision Tokens (InternVIT): CLS tokens for each of the 8 images from layer 2 \\

\end{longtable}


After much experimenting, we used predictions from embeddings marked with an * in Table \ref{tab:vlm_final_embd_perf} in our stacking model. These were selected as they provided the best prediction performance from different embedding classes - Language tokens from LLM layers, Vision tokens from LLM layers and Vision tokens from Visual model layers.


\subsection{Parameters for InternVL feature extraction}
Given the final evaluation videos were to be 2 hours in total length, our goal was to prepare detailed transcripts for the final videos once released. However, due to time constraints we were unable to do so. For the model selection stage we used the SDK simple transcripts. We extracted features for the different videos using the above mentioned embeddings. However, given the simple transcripts we used average of last 3 tokens instead of average of last 10 tokens. We then performed PCA to select 250 features which were used with ridge regression to make predictions.

\begin{table}[ht]
\centering
\caption{Training Details}
\label{tab:subjall_perf}

\begin{tabular}{ll}
\hline
\textbf{Parameter} & \textbf{Value} \\
\hline
Stimulus Window & 2 (Theoretically 1 should be sufficient but our experiments showed an improvement with 2)\\
HRF Delay & 3 \\
Trained On & Friends Seasons 1-6, Bourne Identity, Wolf of Wallstreet, Figures, Life \\
Hardware & 8 x Nvidia RTX A4000 (16 GB RAM) \\
Embeddings Extraction Time & 16 hours for 4 subjects\\
\hline
\end{tabular}
\end{table}


\subsection{Sample comparison base and stimulus-tuned model outputs}
\label{appendix:base-vs-strim-tuned_llama}

\begin{center}
\refstepcounter{roundedboxctr} 
\label{box:suppl_base_x_stim_tuned_llm} 
\begin{roundedbox}[halign=flush left]{Model continuations}

\noindent\textbf{Before:} Stop sending food to our apartment. Why are you all dressed up? You're not the only\textcolor{red}{...}  

\tcblower

\noindent
\begin{minipage}[t]{0.48\textwidth}
\vspace{0pt}            
\raggedright
\setlength{\parskip}{0pt}\setlength{\parindent}{0pt}
\begin{tabular}[t]{@{}p{\linewidth}@{}}

\textbf{Base model continuation:} 

25-year-old woman who gets married and then gets a new haircut. Sheesh. My new hairstyle is fabulous. I love the shorter cut. I love it. You don't. But you know what? This haircut is going to be a little bit short for about two months. And then it's going to get longer and then shorter. It's a little bit like being pregnant. I mean, it's not the same thing at all, but it is. Because for about six weeks I'm just going to have to not be pregnant. And then I'll be pregnant again. And then I'll be pregnant one more time and then I'll have a baby. I have no idea why I said this.

\end{tabular}
\end{minipage}\hfill
\begin{minipage}[t]{0.48\textwidth}
\vspace{0pt}
\raggedright
\setlength{\parskip}{0pt}\setlength{\parindent}{0pt}
\begin{tabular}[t]{@{}p{\linewidth}@{}}

\textbf{Stimulus-tuned model continuation:}
20th century girl who knows how to throw on a great little black dress. Come on. I think we're ready. No, we are not ready. It's not time yet. But I'm hungry. What are we waiting for? \\
It's not time \\
yet. It's not time yet. It's not time yet.  \\
Okay, I will throw this up. I have an extra ticket to the  \\
2000th anniversary of the Statue of Liberty. You got a ticket to that? No, that's where we're going.  \\
Okay, what's the 2000th anniversary? I have no idea, but the people who put it on are very, very fancy. \\
Yeah. Oh, they know people. So, you want to go with us? Yeah, all right. But I'm not dressed. I'm wearing my birthday suit. Okay, you're all set. Hey, you. Hey. It's so beautiful.  \\
Hey, Phoebe, are you  \\
going to this? …
\end{tabular}
\end{minipage}

\end{roundedbox}
\end{center}

\section{slow\_r50 Fine Tuning Implementation Details and Results}
\label{sec:slowr50_details}
\subsection{Architectural Decisions}

We fine tuned slow\_r50 using LoRA adapters. We did experiments on fine tuning a \textit{More Extensive List of Layers} and a \textit{Less Extensive List of Layers} as shown in Box \ref{box:suppl_lora-layers}.

\begin{center}

\refstepcounter{roundedboxctr} 
\label{box:suppl_lora-layers} 
\begin{roundedbox}[halign=flush left]{slow\_r50 layers to fine tune}
\noindent\textbf{More Extensive Layer List} 

blocks.3.res\_blocks.\{0,1,2,3,4,5\}.branch2.conv\_a \\
blocks.3.res\_blocks.\{0,1,2,3,4,5\}.branch2.conv\_b \\
blocks.4.res\_blocks.\{0,1,2\}.branch2.conv\_a \\
blocks.4.res\_blocks.\{0,1,2\}.branch2.conv\_b \\

\tcbline
\noindent\textbf{Less Extensive Layer List}

blocks.3.res\_blocks.\{0,1,2\}.branch2.conv\_c\\
blocks.4.res\_blocks.\{0,1,2\}.branch2.conv\_b\\
blocks.4.res\_blocks.\{0,1,2\}.branch2.conv\_c\\

\end{roundedbox}
\end{center}

\begin{figure}[ht]
    \centering
    
    \begin{subfigure}[b]{0.45\textwidth}
       \centering
        \includegraphics[width=\textwidth]{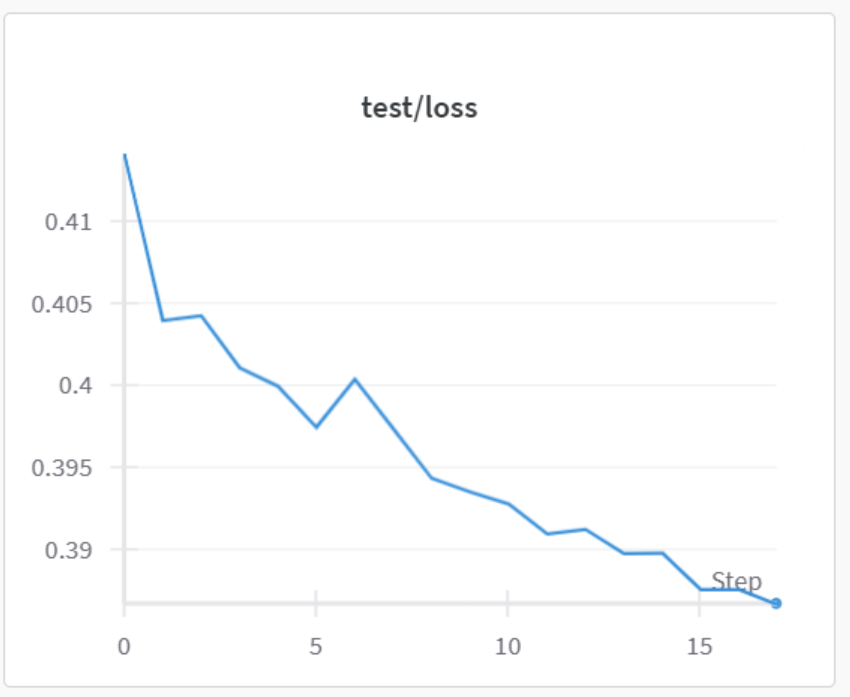}
        \caption{Training Loss}
        \label{fig:slow_loss}
    \end{subfigure}
    \hfill
    \begin{subfigure}[b]{0.45\textwidth}
        \centering
        \includegraphics[width=\textwidth]{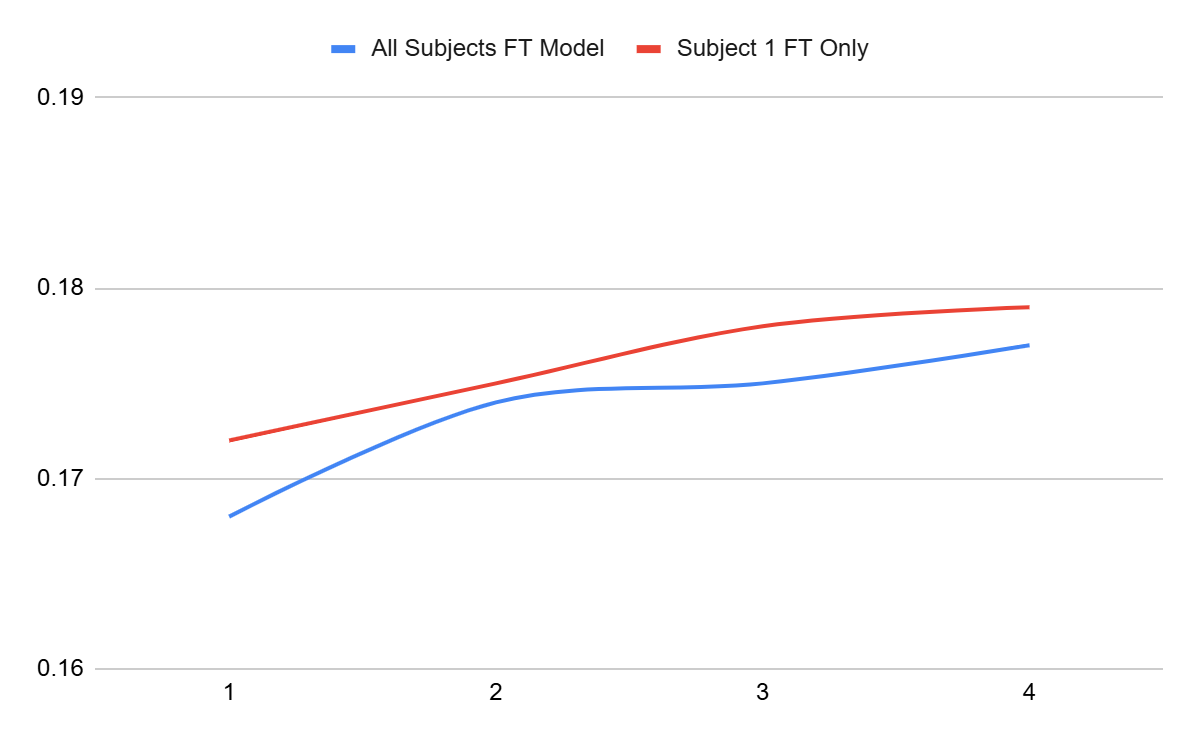}
        \caption{Fine Tuning for all subjects vs 1 subject}
        \label{fig:slow_allsubj}
    \end{subfigure}

    a
    \caption{ a) Training loss by epoch for fine tuning over \textit{Friends} seasons 1 to 5. b) Pearson r score when fine tuning slow\_r50 per subject vs fine tuning for all subjects. }
    \label{fig:slowr50_1}
\end{figure}

The experiments were run by training on \textit{Friends} season 1 to 5 and testing on season 6 for subject 1. We found that the \textit{Less Extensive List of Layers} had slightly better performance and as expected faster training time. We moved ahead with this list. We do believe more performance gains could be had by a more comprehensive parameter selection process - we were unable to do so in interest of time and training resources.

In addition we trained a linear layer between slow\_r50 output (block5 pool) and provided fMRI (functional MRI) predictions as shown in Figure \ref{fig:figure2-complete submission} c. We evaluated the role of training Linear Layer vs fine tuning model parameters. As can be seen in Table \ref{tab:linefreeze_perf}, freezing model parameters and training only Linear Layer resulted in a performance drop.

\begin{center}
\begin{table}[ht]
\centering
\begin{tabular}{lccccc}
\toprule
\textbf{Network} & \textbf{No Fine Tuning (Baseline)} & \textbf{1 Epoch} & \textbf{2 Epochs} & \textbf{3 Epochs}& \textbf{4 Epochs} \\
\midrule
Full Brain & 0.161 & 0.172 & 0.175 & 0.178 & 0.179 \\
Visual Network Only & 0.283 & 0.294 & 0.296 & 0.297 & 0.298 \\
\bottomrule
\end{tabular}
\caption{Pearson correlations for \textit{Friends} season 6 for Subject 1 by number of epochs of fine tuning slow\_r50 model over \textit{Friends} seasons 1-5}
\label{tab:ft_epocn}
\end{table}
\end{center}


\begin{center}
\begin{table}[ht]
\centering
\begin{tabular}{lcc}
\toprule
\textbf{Network} & \textbf{Fine Tune Linear Layer Only} & \textbf{Fine Tune Linear Layer + slow\_r50 Layers} \\
\midrule
Full Brain     & 0.142 & 0.179 \\
Visual Network  Only   & 0.266 & 0.298 \\
\bottomrule
\end{tabular}
\caption{Pearson correlations for Friends Season 6 for Subject 1 for fine-tuning only Linear Layer or both Linear Layer and slow\_r50 \textit{Less Extensive Layer} list. Fine-tuning results after 4 epochs are shown.}
\label{tab:linefreeze_perf}
\end{table}
\end{center}

When fine tuning on \textit{Friends} seasons 1-5 we trained on 80\% of the videos and validated on 20\% of the videos. Our loss would continue to fall even after 16 epochs as can be seen in Figure \ref{fig:slow_loss} and prediction accuracy on season 6 would improve as can be see in Table \ref{tab:ft_epocn}. However, when testing against an out of distribution video like \textit{Life}, we noticed that prediction accuracy would start to decline after 3 epochs - indicating we were overfitting for \textit{Friends} videos. We experimented with fine tuning slow\_r50 with all 4 subjects together - one Linear Layer per subject. Our hope was that this could prevent over fitting. We saw good results(compared to baseline) but as can be seen in Table \ref{tab:subjall_perf} and Figure \ref{fig:slow_allsubj} they could not match performance of fine tuning with just 1 subject at a time. Our final submission for the competition was based on fine tuning a model per subject.

\begin{center}
\begin{table}[ht]
\centering
\begin{tabular}{lcc}
\toprule
\textbf{Epochs} & \textbf{All Subjects} & \textbf{Subject 1 Only} \\
\midrule
1     & 0.162 & 0.172 \\
2   & 0.174 & 0.175 \\
3     & 0.175 & 0.178 \\
4   & 0.177 & 0.179 \\
\bottomrule
\end{tabular}
\caption{Pearson correlations for Friends Season 6 for Subject 1, when model is fine-tuned for all subjects and evaluated for subject 1 vs model fine-tuned for only Subject 1 and evaluated for subject 1}
\label{tab:subjall_perf}
\end{table}
\end{center}

\subsection{Parameters for fine tuning slow\_r50}
When experimenting with Batch Size for training we noticed that it was difficult to converge to a good model when using less than 32. We were limited by available hardware and were not able to experiment with batch sizes over 32. The learning rate we used was also selected after small experiments - unfortunately in interest of time this was not as exhaustive as it could be for an optimal solution. Input features to the vision model were as per baseline provided in competition SDK - 8 equi-spaced frames within a TR at a resolution of 224 x 224px.

\begin{table}[ht]
\centering
\caption{Training Details}

\begin{tabular}{ll}
\hline
\textbf{Parameter} & \textbf{Value} \\
\hline
Stimulus Window & 4 \\
HRF Delay & 3 \\
Trained On & Friends Seasons 1-6, Bourne Identity, Wolf of Wallstreet, Figures, Life \\
Epochs & 3 \\
Learning Rate (LoRA) & Cosine Annealing from $10^{-4}$ to $10^{-7}$ over 10 epochs \\
Learning Rate (Linear Layer) & Cosine Annealing from $10^{-4}$ to $10^{-6}$ over 10 epochs \\
LoRA Weight Decay & $10^{-3}$ \\
Batch Size & 32 \\
Hardware & 4 × Nvidia L40 (94 GB GPU RAM) \\
Training Time & 12 hours for 4 subjects \\
\hline
\end{tabular}
\end{table}

\section{fMRI to fMRI prediction using Hidden Markov Models}
\label{sec:appendix_hmm}

\begin{algorithm}[H]
\caption{fMRI to fMRI prediction using Hidden Markov Models}
\label{alg:hmm}
\begin{algorithmic}[1]
\Require fMRI time series: all parcels $Y$ (for standard Gaussian HMM); predictor parcels $X$ (visual cortex), dependent parcels $Y$ (remaining parcels) (for Gaussian-Linear HMM); number of states $K$
\Ensure Predicted timecourses for all parcels
\vspace{0.5em}
\State \textbf{Preprocessing:}
\Statex \hspace{1em} Standardize timeseries separately for each scanning session
\Statex \hspace{1em} Apply PCA: 
\Statex \hspace{2em} $X \rightarrow$ 10 principal components,
\Statex \hspace{2em} $Y \rightarrow$ 100 principal components

\State \textbf{Model Specification:}
\Statex \hspace{1em} Hidden states $z_t \in \{1, \dots, K\}$
\Statex \hspace{1em} Observation model:
\Statex \hspace{2em} \textbf{Standard Gaussian HMM:} \quad $Y_t \sim \mathcal{N}(\mu_k, \Sigma_k)$
\Statex \hspace{2em} \textbf{Gaussian-Linear HMM (GLHMM):} \quad $Y_t \sim \mathcal{N}(\mu_k + X_t \beta_k, \Sigma_k)$
\Statex \hspace{1em} Temporal evolution:
\Statex \hspace{2em} Initial state probabilities $\pi$
\Statex \hspace{2em} Transition probabilities $A$

\State \textbf{Parameter Estimation:} Estimate $\{\pi, A, \mu_k, \Sigma_k, \beta_k\}$ (where applicable) on training set using variational inference

\State \textbf{Prediction:}
\Statex \hspace{1em} For standard HMM:
\Statex \hspace{2em} Sample from trained model to generate predicted timeseries for all parcels $\hat{Y}$ in test set (fully held-out)
\Statex \hspace{1em} For GLHMM:
\Statex \hspace{2em} (i) Sample from trained model using true predictor timeseries $X$ from test set (not fully held-out) to generate predicted timeseries for worse-predicted parcels $\hat{Y}$ 
\Statex \hspace{2em} (ii) Sample from trained model using predicted timeseries $\hat{X}$ from provider model (test set fully held-out) to generate predicted timeseries for worse-predicted parcels $\hat{Y}$
\Statex \hspace{1em} Back-transform PCs into original dimensionality
\end{algorithmic}
\end{algorithm}

\end{document}